\documentclass[11pt]{article}
\usepackage[utf8]{inputenc}
\usepackage[T1]{fontenc}
\usepackage{authblk}
\usepackage{multicol}
\usepackage{amsmath,amssymb,amsfonts,amsthm}
\usepackage[top=.80in, bottom=.80in, left=.80in, right=.80in]{geometry}
\newcommand{\Keywords}[1]{\par\noindent
{\small{\em Keywords\/}: #1}}

\catcode`,\active

\catcode`\,12

\catcode`,\active

\catcode`\,12

\newcommand{\class}[1]{\par\noindent
{\small{AMS classification scheme numbers\/}: #1}}
\newcommand{\pd}{\partial}

\newcommand{\under}[1]{_{#1}}
\numberwithin{equation}{section}
\title{The Hahn superalgebra and supersymmetric Dunkl oscillator models}
\date{}
\author[1]{Vincent X. Genest\thanks{genestvi@crm.umontreal.ca}}
\author[1]{Jean-Michel Lemay}
\author[1]{Luc Vinet}
\author[2]{Alexei Zhedanov}
\affil[1]{Centre de Recherches Math\'ematiques, Universit\'e de Montr\'eal, C.P. 6128, Succ. Centre-ville, Montr\'eal, QC, Canada, H3C 3J7}
\affil[2]{Donetsk Institute for Physics and Technology, Donetsk 83114, Ukraine}
\begin{document}
\maketitle
\thispagestyle{empty}
\hrule
\begin{abstract}
\noindent
A supersymmetric extension of the Hahn algebra is introduced. This quadratic superalgebra, which we call the Hahn superalgebra, is constructed using the realization provided by the Dunkl oscillator model in the plane, whose Hamiltonian involves reflection operators. In this realization, the reflections act as grading operators and the odd generators are part of the Schwinger-Dunkl algebra, which is a two-parameter extension of the bosonic $\mathfrak{su}(2)$ construction. The even part of the algebra is built from bilinears in the odd generators and satisfy the Hahn algebra supplemented with involutions. A family of supersymmetric Dunkl oscillator models in $n$ dimensions is also considered. The Hamiltonians of these supersymmetric models differ from ordinary Dunkl oscillators by pure reflection terms. In two dimensions, the supersymmetric Dunkl oscillator is seen to have the even part of the Hahn superalgebra as invariance algebra.
\bigskip


\bigskip

\class{81Q60, 17A70, 17A45}
\end{abstract}
\hrule
\section{Introduction}
In the standard approach to one-dimensional supersymmetric quantum mechanics (see section 4.1), the Hamiltonian describing the system is written in terms of supercharge operators which combine spin-like (fermionic) degrees of freedom with the (bosonic) positions \cite{Cooper-2002,Dhoker-1984-09,Dhoker-1985-09,Witten-1981-10}. It has been appreciated that supersymmetry can also be realized in quantum-mechanical systems in which the Hamiltonians do not involve additional degrees of freedom but explicitly contain reflections \cite{Niederle-1999-03,Plyushchay-1996-02,Plyushchay-2000,Post-2011-10}. This approach has been extended to matrix superpotentials in \cite{Nikitin-2011-06} and to $\mathcal{N}$-fold supersymmetry in \cite{Tanaka-2012-07}. Physical models described by Hamiltonians involving reflections have most notably occurred in quantum many-body systems of Calogero-Sutherland type \cite{Calogero-1969-12,Sutherland-1971-11} and their generalizations \cite{Minahan-1993-03}. In these models the constants of motion are expressed using Dunkl operators, which are differential or difference operators involving reflections. These operators also play an important role in the theory of multivariate orthogonal polynomials \cite{Dunkl-2001} and are at the center of Dunkl harmonic analysis (see for example \cite{Li-2013-04}). Finite/discrete models of the oscillator based on algebras with reflections have also been considered recently \cite{Jafarov-2011-05}.

The most simple one-dimensional model involving reflections is the parabose oscillator, also known as the Dunkl oscillator (see Section 2) \cite{Mukunda-1980-10}. This model has an $sl_{-1}(2)$ dynamical algebra generated by the two parabose creation/annihilation operators, the reflection operator and the Hamiltonian \cite{Tsujimoto-2011-10}. The dynamical $sl_{-1}(2)$ can be extended to an $osp(1|2)$ dynamical superalgebra  by incorporating generators obtained from the bilinears in the parabose operators. These bilinear generators commute with the reflections and together with the Hamiltonian satisfy the commutation relations of $\mathfrak{su}(1,1)$, which is the even subalgebra of $osp(1|2)$. Furthermore, the one-dimensional Dunkl oscillator model also admits a supersymmetric extension (Section 4) which can be constructed directly using the results of \cite{Post-2011-10}.

Recently, three of us have introduced in \cite{Genest-2013-04} the Dunkl oscillator model in the plane, which corresponds to the combination of two independent Dunkl oscillators (see Section 2). Using the $sl_{-1}(2)$ dynamical algebra of the one-dimensional constituents, the Dunkl oscillator model in the plane was shown to be superintegrable and the constants of motion were obtained using the Schwinger construction. The algebra formed by these conserved quantities has been determined and found to correspond to a deformation of the Lie algebra $\mathfrak{su}(2)$ by reflection operators. It has been called the Schwinger-Dunkl algebra $sd(2)$. Special cases of $sd(2)$ have also been considered in \cite{Jafarov-2011-05}. The irreducible representations of $sd(2)$ have been investigated in \cite{Genest-2013-09} and generalizations of the two-dimensional model were considered in \cite{Genest-2013-07}. These studies revealed that the Dunkl oscillator model is a showcase for the recently discovered $-1$  orthogonal polynomials of the Bannai-Ito scheme \cite{Genest-2013-09-02,Genest-2013-02-1,Tsujimoto-2012-03,Tsujimoto-2013-03-01, Vinet-2011-01,Vinet-2012-05}. Interestingly, after a gauge transformation, the two-dimensional Dunkl oscillator in the plane can be presented as a singular oscillator model with reflections (see section 2). 

The standard two-dimensional singular oscillator (without reflections) is superintegrable \cite{Winter-1965-06}. Indeed, the $\mathfrak{su}(1,1)$ dynamical algebra of the one-dimensional model can be used to obtain the constants of motion using a Schwinger-like construction \cite{Letourneau-1995-10}. In this case the resulting symmetry algebra is a quadratic algebra known as the Hahn algebra \cite{Galbert-1991-03}. This algebra is a special case of the Racah algebra \cite{Granovskii-1992-07}, which is behind all second-order superintegrable systems in two-dimensions, the Racah problem of $\mathfrak{su}(1,1)$ \cite{Genest-2013-tmp-1} and all hypergeometric orthogonal polynomials of the Askey scheme \cite{Koekoek-2010}.

In this paper, we construct the supersymmetric extension of the Hahn algebra using the realization provided by the Dunkl oscillator model. This superalgebra, that we call the Hahn superalgebra, is obtained by introducing as additional elements bilinears in the $sd(2)$ symmetries of the Dunkl oscillator. In this picture, the Schwinger-Dunkl algebra occurs as a part of a novel quadratic superalgebra whose even part is the Hahn algebra supplemented with reflections. This is akin to the enlargement of the dynamical $sl_{-1}(2)$ algebra of the one-dimensional parabosonic oscillator to $osp(1|2)$. Furthermore, we also define a supersymmetric extension of the Dunkl oscillator model in $n$ dimensions. In this model, the Hamiltonian differs from that of the standard Dunkl oscillator by additional terms in the reflection operators. The supercharges are constructed using a skewed coproduct rule reminiscent of the coproduct rule for $sl_{-1}(2)$. For the two-dimensional case, we show that the supersymmetric model is also superintegrable and that its invariance algebra is the even part of the Hahn superalgebra.

The paper is divided as follows. In the first section, the isotropic Dunkl oscillator model is reviewed. In the second section, the Dunkl oscillator realization of the Hahn superalgebra is constructed. In the third section, the standard approach to supersymmetric quantum mechanics and the alternative approach with reflections are reviewed. The supersymmetry of the Dunkl oscillator in one dimension is surveyed and the supersymmetric Dunkl oscillator model is defined in $n$ dimensions. In the conclusion, we establish that the invariance algebra of the supersymmetric Dunkl oscillator model is the even part of the Hahn superalgebra and we offer an outlook.

\section{The Dunkl oscillator in the plane}
In this section, the essentials of the Dunkl oscillator model in the plane are reviewed. The $sl_{-1}(2)$ dynamical algebra is exhibited and the $sd(2)$ constants of motion and algebra are presented. It is also shown that under an appropriate gauge transformation, the Dunkl oscillator Hamiltonian can be cast in the form of a singular (or caged) oscillator Hamiltonian supplemented with reflection operators.
\subsection{Hamiltonian and dynamical algebra}
The Dunkl oscillator model is defined by the Hamiltonian
\begin{align}
\label{H-Dunkl}
H=-\frac{1}{2}\left[(D_{x_1}^{\mu_{1}})^2+(D_{x_2}^{\mu_{2}})^2\right]+\frac{1}{2}\left[x_1^2+x_2^2\right],
\end{align}
where $D_{x_i}^{\mu_{i}}$, $i=1,2$, stands for the Dunkl derivative
\begin{align}
D_{x_i}^{\mu_{i}}=\pd_{x_i}+\frac{\mu_{i}}{x_i}(1-R_{i}),
\end{align}
and where $R_{i}$ denotes the reflection operator which has the action $R_{i}f(x_i)=f(-x_i)$. The Hamiltonian \eqref{H-Dunkl} corresponds to the combination of two independent one-dimensional Dunkl (or parabose) oscillators
\begin{align}
H=H_{1}+H_{2},
\end{align}
with
\begin{align}
\label{1D-Dunkl}
H_{i}=-\frac{1}{2}(D_{x_i}^{\mu_{i}})^2+\frac{1}{2}x_i^2,\qquad\qquad i=1,2.
\end{align}
The one-dimensional Dunkl oscillator has an $sl_{-1}(2)$ dynamical algebra. Indeed, upon introducing the creation/annihilation operators
\begin{align}
\label{Crea-Anni}
A^{(i)}_{\pm}=\frac{1}{\sqrt{2}}\left(x_{i}\mp D_{x_i}^{\mu_{i}}\right),\quad i=1,2,
\end{align}
and defining $A_0^{(i)}=H_i$, it is directly checked that the following commutation relations hold:
\begin{align}
\label{sl_{-1}(2)}
[A_{0}^{(i)},A_{\pm}^{(i)}]=\pm A_{\pm}^{(i)},\qquad \{A_{+}^{(i)},A_{-}^{(i)}\}=2A_0^{(i)},\qquad \{A_{\pm}^{(i)},R_{i}\}=0,\qquad [A_0^{(i)},R_{i}]=0.
\end{align}
The commutation relations \eqref{sl_{-1}(2)} are the defining relations of the Hopf algebra $sl_{-1}(2)$ \cite{Tsujimoto-2011-10}. This algebra can be extended to an $osp(1|2)$ dynamical superalgebra for the Dunkl oscillator. Indeed, upon introducing the additional generators 
\begin{align}
B_{\pm}^{(i)}=(A_{\pm}^{(i)})^2/2,
\end{align}
it is directly checked that the operators $B_{\pm}^{(i)}$, together with the Hamiltonian $A_{0}^{(i)}$, satisfy the $\mathfrak{su}(1,1)$ commutation relations
\begin{align}
\label{su}
[B_-^{(i)},B_{+}^{(i)}]=A_0^{(i)},\qquad [A_0^{(i)},B_{\pm}^{(i)}]=\pm 2 B_{\pm}^{(i)}.
\end{align}
It follows from the relations \eqref{su} and \eqref{sl_{-1}(2)} that the operators $A_{\pm}^{(i)}$, $A_0^{(i)}$ and $B_{\pm}^{(i)}$ form a basis for the superalgebra $osp(1|2)$ \cite{Kamefuchi-1982}. The even generators $A_{0}^{(i)}$ and $B_{\pm}^{(i)}$ correspond to those which commute with the reflection $R_i$ and the odd generators $A_{\pm}^{(i)}$ are the ones which anticommute with $R_i$.

From the representation theory of $sl_{-1}(2)$, one has that the spectrum $E_{i}$ of the one-dimensional Dunkl oscillator $H_i=A_{0}^{(i)}$ is given by
\begin{align}
E_i=n_i+\mu_i+1/2,\qquad n_i=0,1,\ldots,
\end{align}
provided that $\mu_i>-1/2$. The complete Hamiltonian $H=H_1+H_2$ of the 2D Dunkl oscillator thus has the spectrum
\begin{align}
E_{N}=N+\mu_1+\mu_2+1,\quad N=n_1+n_2=0,1,\ldots,
\end{align}
and exhibits a $(N+1)$-fold degeneracy at energy level $E_N$. The exact solutions of the Hamiltonian \eqref{H-Dunkl} can be obtained by the separation of variables in both Cartesian and polar coordinates. Theses solutions are expressed in terms of generalized Hermite or Jacobi and Laguerre polynomials \cite{Genest-2013-04}.
\subsection{Superintegrability and the Schwinger-Dunkl algebra}
Let us first recall the notion of superintegrability. A quantum system with $d$ degrees of freedom described by a Hamiltonian $\mathcal{H}$ is \emph{maximally superintegrable} if it admits $2d-1$ algebraically independent constants of motion $S_i$, $i=1,\ldots,2d-1$ such that $[S_i,\mathcal{H}]=0$, where one of the symmetries $S_i$ is the Hamiltonian itself. The Dunkl oscillator model in the plane can be shown to be superintegrable by exploiting the $sl_{-1}(2)$ dynamical algebra of the one-dimensional constituents \cite{Genest-2013-04}. Indeed, combining the creation/annihilation operators in the two directions $x_1$, $x_2$, the constants of motion and the symmetry algebra of the system are obtained in a way that parallels the Schwinger construction for the standard 2D isotropic harmonic oscillator \cite{Baym-1969}. Following this construction, we define $J_{\pm}$, $J_0$ as
\begin{align}
\label{Sym}
J_{+}=A_{+}^{(1)}A_{-}^{(2)},\qquad J_{-}=A_{-}^{(1)}A_{+}^{(2)},\qquad J_0=H_1-H_2,
\end{align}
where $A_{\pm}^{(i)}$ is given by \eqref{Crea-Anni} and $H_i$, $i=1,2$, by \eqref{1D-Dunkl}. It is easily checked that $[H,J_{\pm}]=[H,J_0]=0$, and hence that $J_{\pm}$, $J_0$ are constants of motion. A direct computation also shows that they satisfy the commutation relations
\begin{align}
\begin{aligned}
\label{sd(2)}
[J_0,J_{\pm}]&=\pm 2J_{\pm},\qquad \{J_{\pm},R_{i}\}=0,\qquad [J_0,R_{i}]=0,\\
[J_{+},J_{-}]&=J_0+J_0(\mu_1 R_1+\mu_2 R_2)-H(\mu_1R_1-\mu_2R_2),
\end{aligned}
\end{align}
where $i=1,2$ and where $R_i^2=1$. The relations \eqref{sd(2)} are taken to define the Schwinger-Dunkl algebra $sd(2)$. It is easily seen that if one takes $\mu_1=\mu_2=0$ in  \eqref{sd(2)}, one recovers the defining relations of $\mathfrak{su}(2)$. The Schwinger-Dunkl algebra admits the Casimir operator
\begin{align}
\label{Cas}
C=J_0^2+2\{J_{+},J_{-}\}+2(\mu_1R_1+\mu_2R_2)+4\mu_1\mu_2R_1R_2,
\end{align}
which in the realization \eqref{Sym} takes the value
\begin{align}
\label{Value-Cas}
C=H^2-1.
\end{align}
Note that since $P=R_{1}R_{2}$ is also a Casimir operator, there is in fact only one independent reflection operator in the Schwinger-Dunkl algebra. In section 3, the algebra \eqref{sd(2)} shall be recast partly in terms of anticommutators to form the odd part of the Hahn superalgebra.
\subsection{The Dunkl oscillator as a singular oscillator with reflections}
As is easily seen from the expression \eqref{1D-Dunkl}, the one-dimensional Dunkl oscillator Hamiltonian contains a first order term in the derivative. This term can be eliminated by performing the gauge transformation
\begin{align}
\widetilde{H}_i=G(x_i)\,H_i\,G^{-1}(x_i),
\end{align}
with $G(x_i)=|x_i|^{\mu_i}$. Under this gauge transformation, the Hamiltonian \eqref{1D-Dunkl} has the simple expression
\begin{align}
\label{G-Dunkl}
\widetilde{H}_i=\frac{1}{2}\left(-\pd_{x_i}^2+x_{i}^2+\frac{\mu_i^2}{x_i^2}-\frac{\mu_i}{x_i^2}R_{i}\right).
\end{align}
The parabosonic creation/annihilation operators \eqref{Crea-Anni}, which together with $\widetilde{H}_{i}=\widetilde{A}_0^{(i)}$ realize the dynamical $sl_{-1}(2)$ algebra, have the form
\begin{align}
\widetilde{A}^{(i)}_{\pm}=\frac{1}{\sqrt{2}}\left(x_i\mp\pd_{x_i}\pm \frac{\mu_i}{x_i} R_i\right).
\end{align}
This leads to a realization of $osp(1|2)$ (equivalent to \eqref{sl_{-1}(2)}-\eqref{su}) that can be presented as follows:
\begin{subequations}
\begin{gather}
\mathcal{Q}_i=\frac{(\widetilde{A}^{(i)}_--\widetilde{A}^{(i)}_+)R_i}{2}=\frac{1}{\sqrt{2}}\left(\pd_{x_i}R_i-\frac{\mu_i}{x_i}\right),\qquad \mathcal{S}_i=\frac{R_i(\widetilde{A}^{(i)}_++\widetilde{A}^{(i)}_-)}{2i}=i\frac{x_i R_i}{\sqrt{2}},
\\
\mathcal{H}_i=Q^2_i=\frac{1}{2}\left(-\pd_{x_i}^2+\frac{\mu_i^2}{x_i^2}-\frac{\mu_i}{x_i^2}R_{i}\right),\qquad
\mathcal{K}_i=S^2_i=\frac{1}{2}x_i^2,
\\
\mathcal{D}_i=-\frac{1}{2}\{\mathcal{Q}_i,\mathcal{S}_i\}=\frac{i}{2}(x_i\pd_{x_i}+1/2),
\end{gather}
\end{subequations}
with $\mathcal{H}_i$, $\mathcal{D}_i$, $\mathcal{K}_i$ obeying the $\mathfrak{su}(1,1)$ commutation relations corresponding to those of the generators of translations, dilations and special conformal transformations of the line
\begin{align}
[\mathcal{H}_i,\mathcal{D}_i]=i \mathcal{H}_i,\qquad [\mathcal{H}_i,\mathcal{K}_i]=2i\mathcal{D}_i,\qquad [\mathcal{D}_i,\mathcal{K}_i]=i \mathcal{K}_i.
\end{align}
Moreover, it is easily checked that $\{\mathcal{Q}_i,R_i\}=\{\mathcal{S}_i,R_i\}=0$ and that $[\mathcal{H}_i,R_i]=[\mathcal{K}_i,R_i]=[\mathcal{D}_i,R_i]=0$. In this presentation, the Hamiltonian $\widetilde{H}_i$ \eqref{G-Dunkl} is the compact operator
\begin{align}
\widetilde{H}_i=\mathcal{H}_i+\mathcal{K}_i.
\end{align}
Applying the gauge transformation on the two variables $x_1$, $x_2$, it is easily seen from \eqref{G-Dunkl} that the Hamiltonian of the two-dimensional Dunkl oscillator becomes
\begin{align}
\label{Full-Gauge}
\widetilde{H}=-\frac{1}{2}\left(\pd_{x_1}^2+\pd_{x_2}^2\right)+\frac{1}{2}\left(x_1^2+x_2^2+\frac{\mu_1^2}{x_1^2}+\frac{\mu_2^2}{x_2^2}\right)-\frac{\mu_1}{2x_1^2}R_{1}-\frac{\mu_2}{2x_2^2}R_{2}.
\end{align}
As can be observed from \eqref{Full-Gauge}, the Dunkl oscillator Hamiltonian \eqref{H-Dunkl} corresponds, up to a gauge transformation, to a singular (or caged) oscillator supplemented with two reflection operators. Of course, the symmetry algebra and the spectrum of the Hamiltonian are left unchanged under this gauge transformation. Note that the gauge function $G(x_i)$ corresponds to the weight function with respect to which the wavefunctions of the 1D oscillator \eqref{1D-Dunkl} are orthonormal \cite{Genest-2013-04}.
\section{The Hahn superalgebra}
In this section, we introduce the supersymmetric extension of the Hahn algebra. This is done by supplementing the Schwinger-Dunkl algebra with even elements, which commute with the reflections, obtained from the bilinears in some $sd(2)$ generators interpreted to have odd gradings. These bilinears are found to satisfy the defining relations of the Hahn algebra with reflections.
\subsection{Even generators and the Hahn algebra}
Consider the two following generators
\begin{align}
K_{\pm}=J_{\pm}^2.
\end{align}
It is directly checked that these operators commute with the reflections, i.e. $[K_{\pm},R_{i}]=0$, $i=1,2$. Upon using the definitions \eqref{Sym} and the commutation relations \eqref{sd(2)}, a direct computation shows that these generators obey the following relations:
\begin{align}
\begin{aligned}
\label{Cubic}
[J_{0},K_{\pm}]&=\pm 4 K_{\pm},\\
[K_{-},K_{+}]&=J_0^3+J_0(\gamma_1+2\mu_1R_1+2\mu_2R_2)+H(\gamma_2+2\mu_2R_2-2\mu_1R_1),
\end{aligned}
\end{align}
where we have defined
\begin{align*}
\gamma_1=3-H^2-2\mu_1^2-2\mu_2^2,\qquad \gamma_2=2\mu_1^2-2\mu_2^2.
\end{align*}
The cubic algebra \eqref{Cubic} can be cast in the form of the Hahn algebra \cite{Genest-2013-07}. To see this, we introduce new generators $K_0$, $K_1$ defined by
\begin{align}
\label{Def}
K_0=J_0/8,\qquad K_1=(K_{+}+K_{-}+J_0^2/2)/8,
\end{align}
and we take $K_2=[K_0,K_1]$. Upon using the commutation relations \eqref{Cubic}, a simple calculation shows that $K_0$, $K_1$, $K_2$ satisfy the defining relations of the Hahn algebra \cite{Granovskii-1992-07}:
\begin{align}
\begin{aligned}
\label{Dompe}
[K_0,K_1]&=K_2,
\\
[K_1,K_2]&=\{K_0,K_1\}+\frac{1}{8}\,K_0(\gamma_1+2\mu_1R_1+2\mu_2 R_2)+\frac{1}{64}H(\gamma_2+2\mu_2R_2-2\mu_1R_1),
\\
[K_2,K_0]&=K_0^2-\frac{1}{4}\,K_1.
\end{aligned}
\end{align}
\subsection{Presentation of the Hahn superalgebra}
The stage has now been set for the presentation of the supersymmetry algebra of the Dunkl oscillator model. The reflection operators are taken to determine the gradings. The operators $J_{\pm}$, which anticommute with the reflection operators $R_i$, are hence considered as odd elements. The operators $K_0$, $K_1$, $K_2$, which commute with the reflections, are considered as even elements. 

We hence define
\begin{align}
E_0=J_0/8,\qquad E_1=(J_{+}^2+J_{-}^2+J_0^2/2)/8,\qquad E_2=(J_{+}^2-J_{-}^2)/16,
\end{align}
as given by \eqref{Def} and \eqref{Sym} with 
\begin{align}
[E_0,R_{i}]=[E_1,R_{i}]=[E_2,R_{i}]=0,\qquad i=1,2,
\end{align}
and moreover we take
\begin{align}
F_{\pm}=J_{\pm},
\end{align}
as defined by \eqref{Sym}
with 
\begin{align}
\{F_{\pm},R_{i}\}=0,\qquad i=1,2.
\end{align}
With these generators, the symmetry algebra of the Dunkl oscillator in the plane can be cast as a superalgebra. The odd elements satisfy the anticommutation relations:
\begin{subequations}
\label{1}
\begin{align}
\{F_{\pm},F_{\pm}\}&=8E_1\pm 16 E_2-32 E_0^2,\\
\{F_{+},F_{-}\}&=-32 E_0^2-\mu_1 R_1-\mu_2 R_2-2\mu_1\mu_2 R_1R_2+\delta,
\end{align}
\end{subequations}
where $\delta=(H^2-1)/2$. The even/odd relations are given by
\begin{subequations}
\label{2}
\begin{align}
[E_0,F_{\pm}]&=\pm F_{\pm}/4,\\
[E_1,F_{\pm}]&=\{E_0,F_{\pm}\}-\{E_0,F_{\mp}\}-F_{\mp}(\mu_1R_1+\mu_2R_2)/4,\\
[E_2,F_{\pm}]&=\{E_0,F_{\mp}\}/2\pm F_{\mp}(\mu_1R_1+\mu_2R_2)/8,
\end{align}
\end{subequations}
and the even generators satisfy the Hahn algebra relations
\begin{subequations}
\label{3}
\begin{align}
[E_0,E_1]&=E_2,
\\
[E_1,E_2]&=\{E_0,E_2\}+\frac{1}{4}E_0(\omega_1+\mu_1R_1+\mu_2R_2)+\frac{1}{32}H(\omega_2+\mu_2R_2-\mu_1 R_1),
\\
[E_2,E_0]&=E_0^2-E_1/4,
\end{align}
\end{subequations}
where 
\begin{align*}
\omega_1=3/2-H^2/2-\mu_1^2-\mu_2^2\qquad \omega_2=\mu_1^2-\mu_2^2.
\end{align*}
The quadratic superalgebra with the defining relations \eqref{1}, \eqref{2} and \eqref{3} corresponds to the supersymmetric extension of the Hahn algebra. The even part with generators $E_0$, $E_1$ and $E_2$ satisfy the commutation relations of the Hahn algebra and the odd generators $F_{\pm}$, together with $E_0$, are part of the Schwinger-Dunkl algebra. The Casimir operator \eqref{Cas} of the $sd(2)$ algebra is the Casimir operator for the Hahn superalgebra. 
\section{Supersymmetric extension of the Dunkl oscillator}
In this section, we define the supersymmetric extension in $n$ dimensions. We begin by reviewing the standard approach to one-dimensional supersymmetric quantum mechanics and the approach with reflections proposed in \cite{Post-2011-10}. We then review the supersymmetry of the one-dimensional Dunkl oscillator. The model is finally extended to $n$ dimensions and the supercharges are given explicitly.
\subsection{Supersymmetric quantum mechanics with reflections}
Let us first recall the definition of a one-dimensional supersymmetric system. A physical system described by a Hamiltonian $H$ is said to be supersymmetric if it admits supercharges $Q$, $Q^{\dagger}$ such that the following relations are satisfied:
\begin{align}
H=\frac{1}{2}\{Q,Q^{\dagger}\},\qquad [Q,H]=0,\qquad [Q^{\dagger},H]=0.
\end{align}
\subsubsection{Standard approach} 
In the most simple setting of one-dimensional quantum mechanics, the conditions for supersymmetry can be realized by taking
\begin{align}
\label{Super-1}
Q=i(\pd_{x}+U(x))b,
\end{align}
where $U(x)$ is the superpotential and where $b$, $b^{\dagger}$ are the fermionic creation/annihilation operators which satisfy
\begin{align}
\{b,b^{\dagger}\}=1,\qquad b^2=(b^{\dagger})^2=0,
\end{align}
and which are represented by the $2\times 2$ matrices
\begin{align}
b=
\begin{pmatrix}
0 & 1\\
0 & 0
\end{pmatrix},
\qquad
b^{\dagger}=
\begin{pmatrix}
0& 0\\
1 & 0
\end{pmatrix}.
\end{align}
Using the supercharge \eqref{Super-1}, it is directly found that the corresponding Hamiltonian is given by
\begin{align}
\label{Rel-2}
H=\frac{1}{2}\{Q,Q^{\dagger}\}=\frac{1}{2}\Big(-\pd_{x}^2+U(x)^2+U'(x)\sigma_3\Big),
\end{align}
where $\sigma_{3}=\mathrm{diag}(1,-1)$. The relation \eqref{Rel-2} defines a supersymmetric Hamiltonian in one dimension that admits two supercharges $Q$ and $Q^{\dagger}$. This corresponds to $\mathcal{N}=1$ supersymmetry. In the case where there is only one Hermitian supercharge such that $Q=Q^{\dagger}$ and $H=Q^2$, one speaks of $\mathcal{N}=1/2$ or chiral supersymmetry.
\subsubsection{Approach with reflections}
In one-dimension, chiral supersymmetry can be realized with reflection operators. For this purpose, consider the following Dunkl operator:
\begin{align}
\label{Super-R}
Q=\frac{1}{\sqrt{2}}\Big(\pd_{x}+V(x)\Big)R+\frac{1}{\sqrt{2}}W(x),
\end{align}
where $Rf(x)=f(-x)$ is the reflection operator and where the following parity conditions hold on the potentials $V(x)$, $W(x)$:
\begin{align}
V(-x)=V(x),\qquad W(-x)=-W(x).
\end{align}
The operator $R$ is clearly self-adjoint with respect to the standard inner product of functions over symmetric domains, whence it follows that $Q^{\dagger}=Q$. Upon calculating $Q^{2}$, one finds a supersymmetric Hamiltonian with the expression
\begin{align}
H=Q^{2}=\frac{1}{2}\Big(-\pd_{x}^{2}+V^2(x)+W^2(x)+V'(x)-W'(x)R\Big).
\end{align}
Even though the Dunkl supercharges lead to Hamiltonians which can be in certain cases presented in superficially similar forms to those obtain with \eqref{Super-1}, it should be stressed that the two approaches, the standard one and the one with reflections, are genuinely different \cite{Post-2011-10}.
\subsection{Supersymmetry of the 1D Dunkl oscillator}
The approach to supersymmetry with reflections can be used to define a supersymmetric extension of the one-dimensional Dunkl oscillator. Indeed, upon taking
\begin{align}
V(x)=0,\qquad W(x)=x-\frac{\mu}{x},
\end{align}
in \eqref{Super-R}, it is easily checked that one obtains the supersymmetric Hamiltonian
\begin{align}
\label{SUSY-H}
H=Q^{2}=\frac{1}{2}\left(-\pd_x^2+x^2+\frac{\mu^2}{x^2}-\frac{\mu}{x^2}R\right)-\frac{1}{2}R-\mu,
\end{align}
where the Hermitian supercharge $Q$ has the expression
\begin{align}
Q=\frac{1}{\sqrt{2}}\left(\pd_{x}R+x-\frac{\mu}{x}\right).
\end{align}
The supersymmetric Hamiltonian \eqref{SUSY-H} differs from that of the (gauge rotated) one-dimensional Dunkl oscillator \eqref{G-Dunkl} only by a constant and a non-trivial reflection term. Since however these two contributions commute with the original Hamiltonian \eqref{G-Dunkl}, the exact solutions of \eqref{SUSY-H} can be obtained directly from the solutions of the Dunkl oscillator.
\subsection{Supersymmetric Dunkl oscillator in $n$ dimensions}
It is possible to define a supersymmetric model of the Dunkl oscillator in $n$ dimensions. To illustrate the idea, let us consider first the $n=2$ case. Introduce the supercharge
\begin{align}
\label{Super-2}
Q=Q_{1}R_{2}+Q_{2},
\end{align}
where $R_if(x_i)=f(-x_i)$ and where $Q_1$ and $Q_2$ are the supercharges corresponding to the one-dimensional supersymmetric model
\begin{align*}
Q_i=\frac{1}{\sqrt{2}}\big(\pd_{x_i}R_{i}+x_i-\frac{\mu_i}{x_i}\big).
\end{align*}
A direct computation then shows that
\begin{align}
\label{2D-H}
H=Q^{2}=H_{1}+H_{2},
\end{align}
where $H_i$ is the one-dimensional supersymmetric Dunkl Hamiltonian
\begin{align*}
H_i=\frac{1}{2}\left(-\pd_{x_i}^2+x_i^2+\frac{\mu^2_i}{x_i^2}-\frac{\mu_i}{x_i^2}R_{i}\right)-\frac{1}{2}R_{i}-\mu_i,
\end{align*}
The form of $Q$ in \eqref{Super-2} is reminiscent of the coproduct rule for $sl_{-1}(2)$ \cite{Tsujimoto-2011-10}. It is not hard to see that this extends to any dimension. Indeed, if one takes as supercharge
\begin{align}
Q=\sum_{i=1}^{n}Q_{i}R_{i+1}\cdots R_{n},
\end{align}
then one has the associated supersymmetric Hamiltonian with reflections
\begin{align}
H=Q^2=\sum_{i=1}^{n}Q_{i}^2,
\end{align}
which corresponds to a $n$-dimensional supersymmetric extension of the Dunkl oscillator Hamiltonian.
\section{Conclusion}
In this paper, we have used the Dunkl oscillator model in the plane to construct a quadratic superalgebra which corresponds to the supersymmetric extension of the Hahn algebra. We have also defined a supersymmetric extension of the Dunkl oscillator model in two dimensions. The corresponding supersymmetric Hamiltonians differ from that of the standard Dunkl oscillator by pure reflection terms.

In view of this fact, it is easy to see that the even part of the Hahn superalgebra, which corresponds to the Hahn algebra with reflections, is also the symmetry algebra of the supersymmetric Dunkl oscillator model in 2D as described by the Hamiltonian \eqref{2D-H}. Indeed, the additional generators $K_{\pm}=J_{\pm}^2$ built from the squares of the Schwinger-Dunkl algebra elements, commute with both the 2D Dunkl oscillator and the reflections. Hence, these generators also commute with the supersymmetric Hamiltonian \eqref{2D-H} and are thus constants of motion. It follows that in addition to being supersymmetric, the model defined by \eqref{2D-H} is also superintegrable, with the Hahn algebra with reflection as invariance algebra.

Recently, it was shown that all second-order superintegrable systems in two-dimensions can be obtained from a single system: the generic system on the two sphere \cite{Kalnins-2013-05}. It was also shown that this model is equivalent to the Racah problem for $\mathfrak{su}(1,1)$, which is governed by the Racah algebra \cite{Genest-2013-tmp-1}. Given that the Dunkl oscillator is essentially a singular oscillator model with reflections and that the Racah problem for $sl_{-1}(2)$ has already been solved and shown to be governed by the Bannai-Ito algebra \cite{Genest-2012}, it would be of interest to study the supersymmetric extension of the Racah algebra using a Dunkl oscillator model on the 2-sphere.

\section*{Acknowledgements}
V.X.G holds an Alexander-Graham-Bell fellowship from the Natural Science and Engineering Research Council of Canada (NSERC). The research of L.V. is supported in part by NSERC.
\begin{multicols}{2}
\footnotesize

\end{multicols}
\end{document}